\documentclass[superscriptaddress,aps,preprint,floatfix]{revtex4-1}
\usepackage{color}
\usepackage{float}
\usepackage{amsmath}
\usepackage{graphicx}
\usepackage{amssymb}
\usepackage{esint}
\usepackage{setspace}

\begin{document}

\title{Controlled catch and release of microwave photon states}

\author{Yi Yin}
\affiliation{Department of Physics, University of California, Santa Barbara, CA 93106, USA}
\author{Yu Chen}
\affiliation{Department of Physics, University of California, Santa Barbara, CA 93106, USA}
\author{Daniel Sank}
\affiliation{Department of Physics, University of California, Santa Barbara, CA 93106, USA}
\author{P. J. J. O'Malley}
\affiliation{Department of Physics, University of California, Santa Barbara, CA 93106, USA}
\author{T. C. White}
\affiliation{Department of Physics, University of California, Santa Barbara, CA 93106, USA}
\author{R. Barends}
\affiliation{Department of Physics, University of California, Santa Barbara, CA 93106, USA}
\author{J. Kelly}
\affiliation{Department of Physics, University of California, Santa Barbara, CA 93106, USA}
\author{Erik Lucero}
\affiliation{Department of Physics, University of California, Santa Barbara, CA 93106, USA}
\author{Matteo Mariantoni}
\affiliation{Department of Physics, University of California, Santa Barbara, CA 93106, USA}
\affiliation{California NanoSystems Institute, University of California, Santa Barbara, CA 93106, USA}
\author{A. Megrant}
\affiliation{Department of Materials, University of California, Santa Barbara, CA 93106, USA}
\author{C. Neill}
\affiliation{Department of Physics, University of California, Santa Barbara, CA 93106, USA}
\author{A. Vainsencher}
\affiliation{Department of Physics, University of California, Santa Barbara, CA 93106, USA}
\author{J. Wenner}
\affiliation{Department of Physics, University of California, Santa Barbara, CA 93106, USA}
\author{Alexander N. Korotkov}
\affiliation{Department of Electrical Engineering, University of California, Riverside CA 92521, USA}
\author{A. N. Cleland}
\affiliation{Department of Physics, University of California, Santa Barbara, CA 93106, USA}
\affiliation{California NanoSystems Institute, University of California, Santa Barbara, CA 93106, USA}
\author{John M. Martinis}
\email{martinis@physics.ucsb.edu}
\affiliation{Department of Physics, University of California, Santa Barbara, CA 93106, USA}
\affiliation{California NanoSystems Institute, University of California, Santa Barbara, CA 93106, USA}

\clearpage

\begin{abstract}
The quantum behavior of superconducting qubits coupled to resonators is very similar to that of atoms in optical cavities \cite{kimble:1998, haroche:2006}, in which the resonant cavity confines photons and promotes strong light-matter interactions. The cavity end-mirrors determine the performance of the coupled system, with higher mirror reflectivity yielding better quantum coherence, but higher mirror transparency giving improved measurement and control, forcing a compromise. An alternative is to control the mirror transparency, enabling switching between long photon lifetime during quantum interactions and large signal strength when performing measurements. Here we demonstrate the superconducting analogue, using a quantum system comprising a resonator and a qubit, with variable coupling to a measurement transmission line. The coupling can be adjusted through zero to a photon emission rate 1,000 times the intrinsic photon decay rate. We use this system to control photons in coherent states as well as in non-classical Fock states, and dynamically shape the waveform of released photons. This has direct applications to circuit quantum electrodynamics \cite{schoelkopf:2008}, and may enable high-fidelity quantum state transfer between distant qubits, for which precisely-controlled waveform shaping is a critical and non-trivial requirement \cite{cirac:1997, korotkov:2011}.
\end{abstract}

\maketitle

Superconducting resonators play a central role in quantum information technology. Applications include the synthetic generation and storage of photon quantum states \cite{hofheinz:2008, hofheinz:2009, wang:2009b}, quantum memories for quantum computation \cite{mariantoni:2011}, and dispersive measurements of superconducting qubits \cite{blais:2004, wallraff:2005} as well as defects in diamond \cite{kubo:2010, schuster:2010}. Resonators with low internal losses are typically desirable, but the resonator's coupling strength to the quantum system and to its measurement apparatus is application-dependent. When coupling a resonator to a qubit, either for a quantum memory or in a circuit quantum electrodynamics experiment, strong coupling to the qubit improves information transfer but also increases dephasing. When reading out a qubit, coupling the resonator strongly to its measurement apparatus increases the measurement bandwidth and signal but in addition increases dissipation \cite{mallet:2009}. Resonator designs therefore involve compromises between the competing needs for both strong and weak coupling \cite{johnson:2010, leek:2010}. A resonator with a variable coupling would provide a significant improvement: If used to measure a qubit, the coupling to the measurement apparatus could be turned off except during resonator readout, when the coupling could be made large. When coupling two qubits through a resonator, the coupling could be turned on and off as needed \cite{allman:2010, srinivasan:2011}, yielding higher fidelity gates \cite{cirac:1997, korotkov:2011}.

Here we employ an externally-controlled variable inductance \cite{bialczak:2011} to modulate the coupling of a resonator to a transmission line, creating the microwave equivalent of a Fabry-Perot cavity with a variable-transparency mirror. The resonator also has fixed coupling to a superconducting phase qubit. We demonstrate the time-controlled release of single-photon Fock and superposition states, thus generating a ``flying qubit'' \cite{boyzigit:2011, houck:2007, divincenzo:2000}.  We also perform timed capture and release of few-photon coherent states, and use the variable coupling to transmit and release photons with arbitrary waveforms \cite{lukin:2003, keller:2004}. This new capability promises numerous applications in high-fidelity quantum computing and communication.

\begin{figure}[H]
\begin{center}
\includegraphics[scale=0.5]{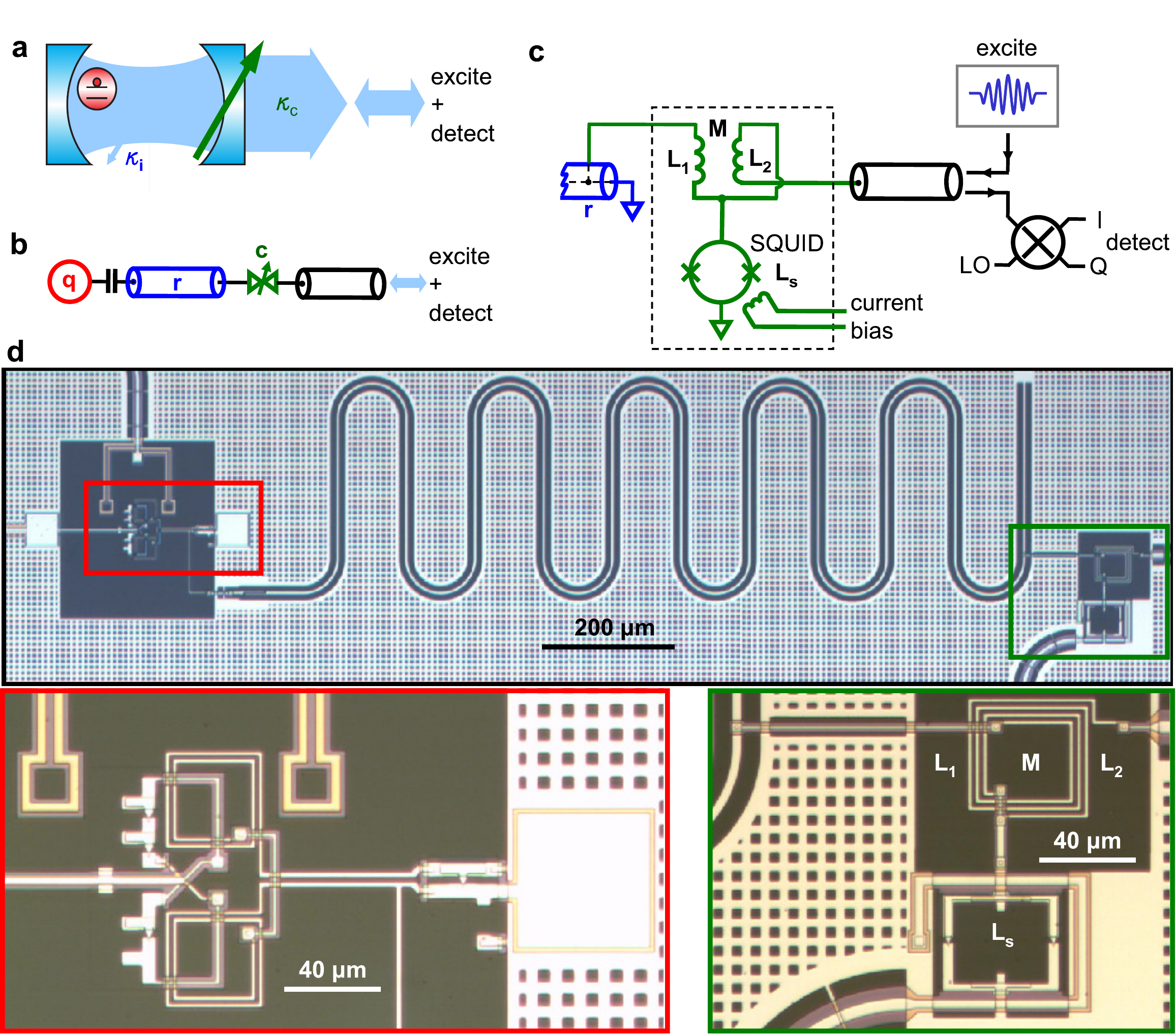}
\caption{Experimental setup.
\textbf{a}, Schematic for cavity quantum electrodynamics with a variable transparency mirror, where $\kappa_c$ is the decay rate through the variable mirror and $\kappa_i$ the intrinsic photon decay rate. The cavity, two-level atom and variable transparency mirror represent the resonator, qubit and variable coupler, respectively.
\textbf{b}, Experimental schematic: The left end of the resonator ($r$) is capacitively-coupled to a superconducting phase qubit ($q$) with coupling $g/2\pi\simeq12\ \text{MHz}$. The resonator is 4.8 mm long with resonant frequency $f_r\simeq6.57$ GHz and the qubit is tunable over the range from $\sim6-7$ GHz. The resonator is connected to a transmission line through a variable coupler ($c$).
\textbf{c}, The variable coupler is a transformer connected $\lambda/60 \approx 0.3$ mm from the grounded end of the resonator $r$, and comprises two fixed inductors $L_1=L_2=480$ pH with a negative mutual inductance $M=-138$ pH, and a SQUID with tunable inductance $L_s(\Phi)$. The SQUID inductance is modulated using the flux from a current bias line through a $50$ pH mutual inductance, with a flux tuning of $\approx \pm\Phi_0/2$ with $\pm 25~\mu$A of bias current. Microwave excitation signals drive the resonator through the transmission line, and signals from the resonator are amplified and demodulated using a mixer driven by a local oscillator (LO). The demodulated $I$ and $Q$ signals oscillate at the LO sideband frequency.
\textbf{d}, Micrograph of device (top), with details of phase qubit (bottom left) and variable coupler (bottom right).\label{fig.expsetup}}
\end{center}
\end{figure}

The schematic in Fig.\,\ref{fig.expsetup}a displays a Fabry-Perot cavity, which represents the resonator, with a tunable transparency mirror to represent the variable coupler. A two-level atom plays the role of the qubit. In the actual experiment (Fig.\,\ref{fig.expsetup}b, c and d), the resonator ($r$) is a quarter-wavelength ($\lambda/4$) coplanar waveguide resonator, with one end coupled to a superconducting phase qubit ($q$) and the other end shorted to ground. Close to the grounded end (a distance $\sim \lambda/60$ away), the resonator is connected to a variable coupler ($c$), which modulates the inductive coupling to a microwave transmission line. The variable coupler is controlled using a bias current, which flux-modulates the inductance of a superconducting quantum interference device (SQUID) embedded in a mutual inductance circuit. The resonator-transmission line coupling $\kappa_c$ can vary from zero to a maximum emission rate $\kappa_{\rm max} \approx 1/(5~{\rm ns})$, over a time scale of a few nanoseconds. The resonator frequency is $f_r\simeq6.57$~GHz, and the phase qubit has a ground to excited state ($|g\rangle \leftrightarrow |e\rangle$) transition frequency tunable from $\sim6$ to 7 GHz \cite{hofheinz:2009, bialczak:2011, yin:2012}. The qubit-resonator coupling is $g/2\pi\simeq12$~MHz, measured using swap spectroscopy \cite{mariantoni:2011}. The qubit-resonator interaction is controlled by tuning the qubit frequency, and is effectively turned off by setting the qubit frequency to its idle point, $400$ MHz below the resonator frequency \cite{hofheinz:2009, bialczak:2011, yin:2012}.

\begin{figure}[H]
\begin{center}
\includegraphics[scale=0.5]{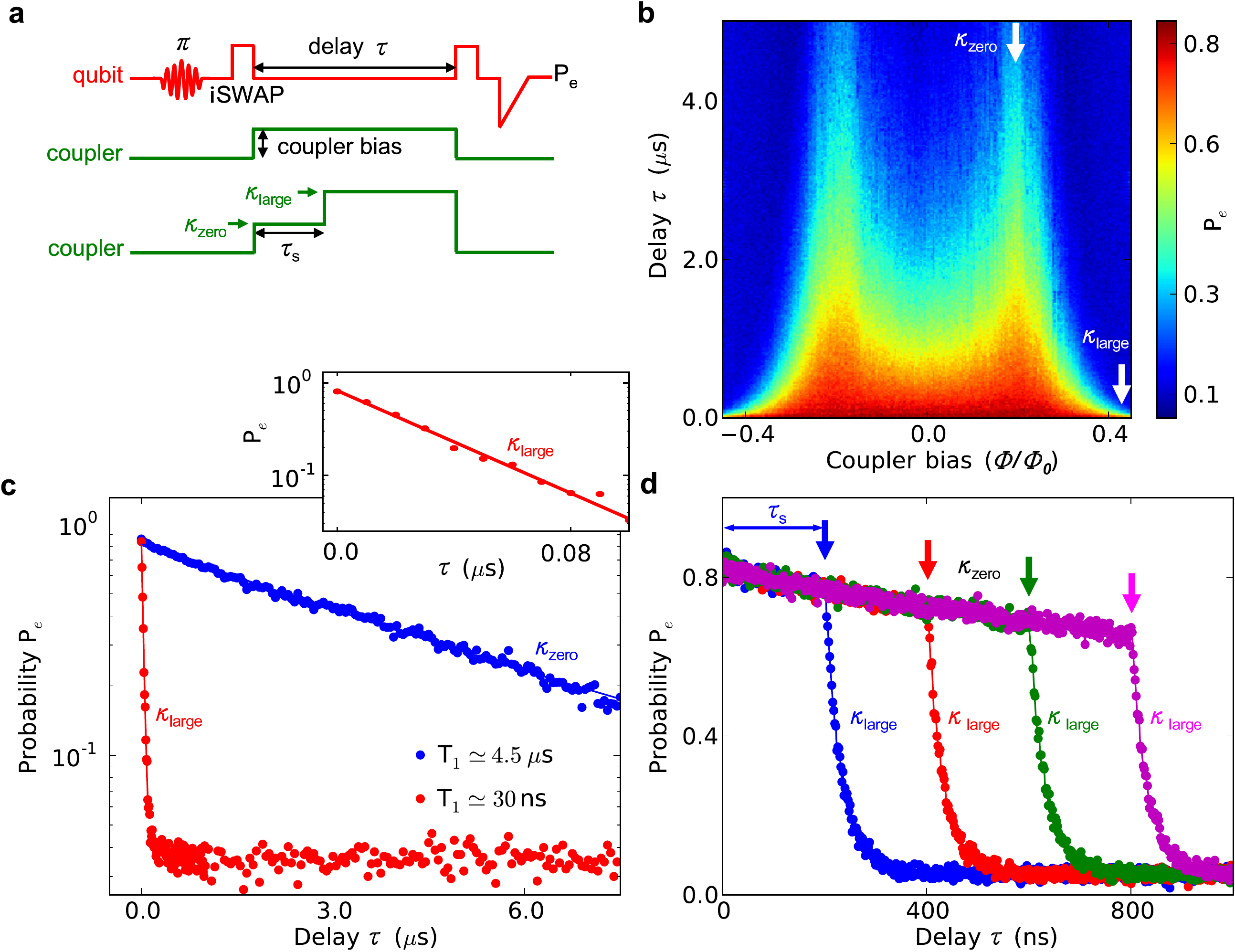}%U:/Yi/Latex/DynamicQ/paperfigure/
\caption{Characterization of the variable coupler with Fock state $|1\rangle$.
\textbf{a}, Pulse sequence. Top: The qubit (red), de-tuned $400$ MHz below the resonator, is excited to $|e\rangle$ by a $\pi$ pulse, then tuned into resonance with the resonator to perform an iSWAP, generating a one-photon Fock state. After a delay time $\tau$, a second pulse transfers the resonator state back to the qubit, and the qubit is measured using a projective single-shot measurement. Averaging 600 times yields the qubit excited state probability $P_e$. Middle: Coupler bias (green) used for \textbf{b} and \textbf{c}, starting with zero current bias followed by a rectangular pulse with variable amplitude setting the coupler strength during the delay period. Bottom: Coupler bias (green) for \textbf{d}, where coupling is switched from zero current to zero coupling ($\kappa_{\rm zero}$) to $\kappa_{\rm large}$ after a time $\tau_s$.
\textbf{b}, Excited probability $P_e$ (color scale) versus delay $\tau$ (vertical axis) and coupler bias in flux units (horizontal axis). $P_e$ decays exponentially with $\tau$ due to combined photon emission and intrinsic loss, with resonator lifetime $T_1$ set by coupler bias.
\textbf{c}, Vertical line-cuts of \textbf{b} display exponential decay of $P_e$, which gives resonator lifetime $T_1$. Resonator lifetime for zero coupling ($\kappa_{\rm zero}$) is intrinsic $T_{\rm 1,i}=4.5\ \mu\text{s}$, while for $\kappa_{\rm large}$ it is reduced to $T_1=30$ ns. Inset shows $P_e$ for short times for $\kappa_{\rm large}$.
\textbf{d}, $P_e$ versus delay $\tau$ for coupling strength switching from zero to $\kappa_{\rm large}$. Decay rate switches from intrinsic lifetime ($4.5~\mu$s) to 30 ns, with transition taking $\sim 2$ ns. Blue, red, green and purple lines correspond to switching delays $\tau_s$ of 200, 400, 600 and 800 ns, respectively. \label{fig.onephoton}}
\end{center}
\end{figure}

The resonator and variable coupler were characterized by measuring the decay of a one-photon Fock state stored in the resonator. When the resonator is weakly coupled to the transmission line, the photon decays due to internal resonator losses, while when strongly coupled, the photon is emitted into the transmission line. The pulse sequence is shown in Fig.\,\ref{fig.onephoton}a, where the qubit was excited from $|g\rangle$ to $|e\rangle$, and the excitation then swapped to the resonator, creating a one-photon Fock state \cite{hofheinz:2009}. The coupler bias current was then adjusted, and after a delay time $\tau$, the residual excitation swapped back to the qubit, and the qubit measured.

Figure\,\ref{fig.onephoton}b displays the probability $P_e$ of measuring the qubit in $|e\rangle$ as a function of delay $\tau$ and the variable coupler current bias. $P_e$ decays exponentially with time $\tau$, with the decay rate varying strongly with coupler bias. Two line cuts are shown in Fig.\,\ref{fig.onephoton}c, with exponential fits yielding the resonator lifetime $T_1$. For zero coupling, as determined from maximizing $T_1$ with respect to coupler bias, we find the intrinsic $T_{1,i} \approx 4.5~\mu$s, in agreement with resonator loss measurements, while for coupling $\kappa_{\rm large}$ the lifetime is reduced to $T_{1} \approx 30$ ns. The resonator inverse lifetime $1/T_1$ is the sum of the intrinsic decay rate $1/T_{\rm 1,i}$ and the coupler emission rate $\kappa_c$, so $\kappa_c = 1/T_1 - 1/T_{\rm 1,i}$. The coupling dependence on current bias in Fig.\,\ref{fig.onephoton}b is in good agreement with calculations (Supplementary Information).

We demonstrated dynamic control by changing the coupling during the delay period, as shown in Fig.\,\ref{fig.onephoton}d. We started with the coupling set to zero, and after a delay $\tau_s$ switched the coupling to $\kappa_{\rm large} \simeq 1/(30~{\rm ns})$. The reduction in the photon lifetime after the switch is clearly visible. The coupler switching speed was limited by the $\sim 2$ ns rise time of the coupler bias, roughly 2,000 times shorter than $T_{1,i}$.

\begin{figure}[H]
\begin{center}
\includegraphics[scale=0.5]{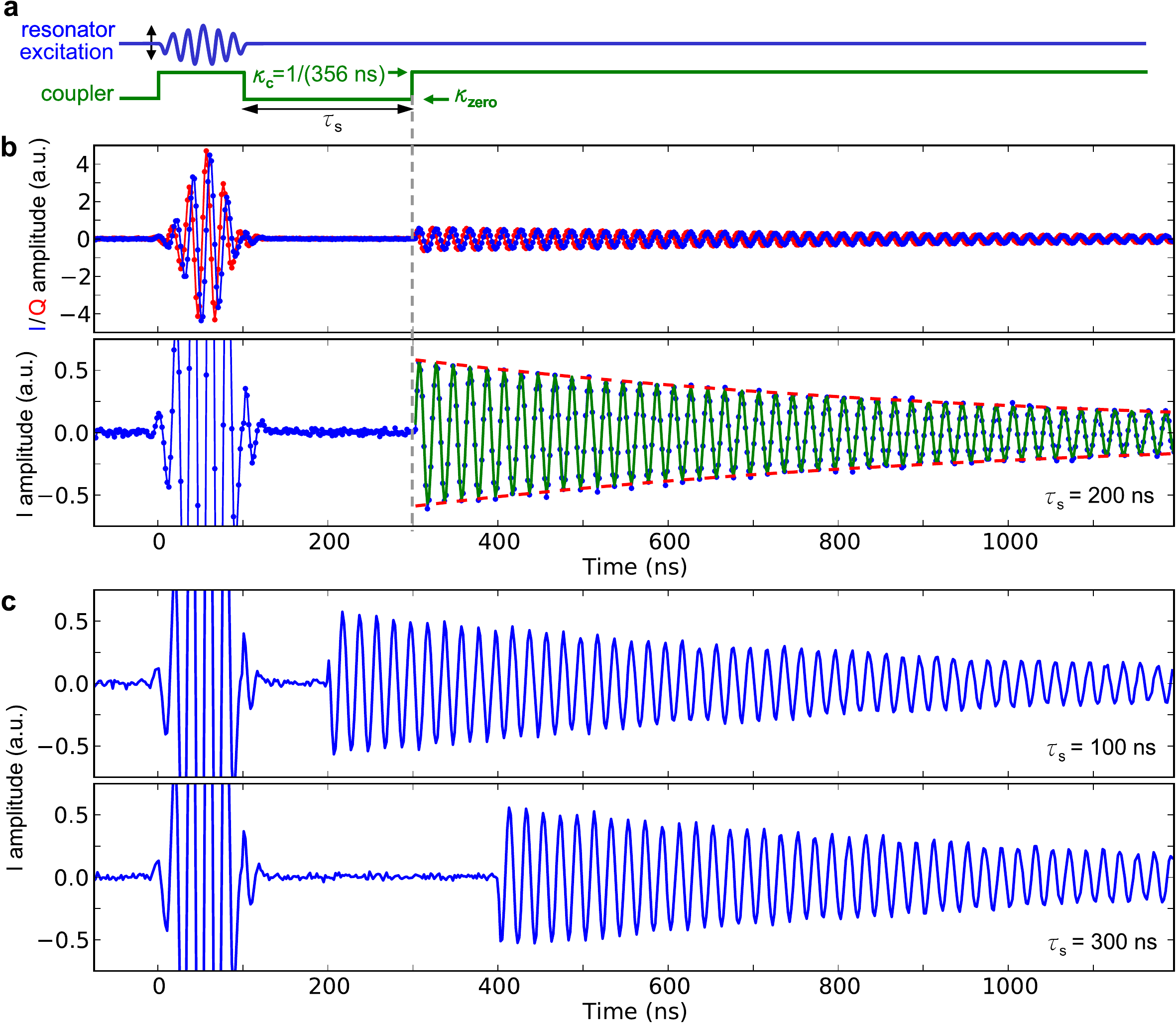}
\caption{Catch and release of photons in coherent states. All traces averaged $10^5$ times.
\textbf{a}, Pulse sequence. Resonator (blue) is driven by on-resonance, 100 ns Gaussian excitation pulse (50 ns full-width at half-maximum (FWHM)), and coupler (green) tuned to $\kappa_{c}\simeq 1/(356~{\rm ns})$. Gaussian pulse is calibrated so resonator catches a coherent state with $\langle n \rangle = 10$ photons. Photons are stored for time $\tau_s$ at zero coupling, then released with $\kappa_{c} = 1/(356~{\rm ns})$.
\textbf{b}, Top sub-panel: Demodulated $I$ and $Q$ quadrature signals for $\tau_s = 200$ ns, with 50 MHz sideband oscillations and a relative $\pi/2$ phase shift (lines are guides to the eye). Signals include reflected part of excitation pulse, followed by release signal after delay $\tau_s$, comprising a sudden onset with exponential decay. Bottom sub-panel: $I$ on expanded scale, with sinusoidal fit (green line), comprising sideband oscillations with exponential decay envelope (dashed red line, time constant $T_d = 706$ ns). The fixed phase with $10^5$ averages indicates phase coherence of photon release.
\textbf{c}, $I$ quadrature for trapping delays $\tau_s = 100$ ns and 300 ns, showing excitation pulse and a delayed photon release, with a delay-dependent phase shift.   \label{fig.coherent}}
\end{center}
\end{figure}

This measurement does not distinguish between incoherent decay and the expected phase-coherent release of the photon. We therefore also used heterodyne detection, with the resonator ``catching'' and then ``releasing'' photons in coherent states. Figure\,\ref{fig.coherent}a displays the pulse sequence: With the coupler set to an intermediate coupling $\kappa_c = 1/(356~{\rm ns})$, we excited the resonator with a 100 ns Gaussian pulse from the microwave source, with the pulse calibrated to trap $\langle n \rangle = 10$ photons (see Supplementary Information). The coupling was then set to zero, trapping the photons for a storage time $\tau_s$, then set back to $\kappa_c = 1/(356~{\rm ns})$, releasing the photons for heterodyne detection.

Figure\,\ref{fig.coherent}b shows the heterodyne-detected signal in the $I$ (real) and $Q$ (imaginary) quadratures in the time domain. During the Gaussian excitation pulse, the signal comprised the reflected component of the excitation. No signal was detected during the subsequent $\tau_s = 200$ ns storage time with the coupler turned off. A sharp onset followed by an exponentially-decaying envelope appeared when the coupler was turned back on, releasing the photons. The signal envelope has a decay time $T_d = 706$ ns, in close agreement with the expected value $2/\kappa_c = 712$ ns. The amplitude oscillations are from a 50 MHz mixer sideband signal, and the $I$ and $Q$ quadratures have a relative $\pi/2$ offset, as expected.   As the output traces were averaged $10^5$ times, the presence of oscillations indicates that the output represents coherent photon release, with a fixed output phase relative to the input.

Figure\,\ref{fig.coherent}c displays the $I$ quadratures using storage times $\tau_s = 100$ and 300 ns. These are identical during the state-generating Gaussian pulse, but during the release the oscillation phase depends on the storage time $\tau_s$, scaling as $\sim 0.81 \pi (\tau_s/100~{\rm ns})$. This phase accumulation is as expected from the small tuning of the resonator frequency $f_r$ with coupler bias (see Supplementary Information), further demonstrating the coherence of the release.

We also calculated the radiated energy $\int_{t_r}^{t_c} (I^2(t)+Q^2(t)) dt$, integrating the signal power from the photon release time $t_r$ to a cutoff $t_c = t_r+3 T_d$. We find that the released energy for $\tau_s =300$ ns is 4\% lower than for $\tau_s = 100$ ns, in agreement with the expected intrinsic resonator loss.

\begin{figure}[H]
\begin{center}
\includegraphics[scale=0.5]{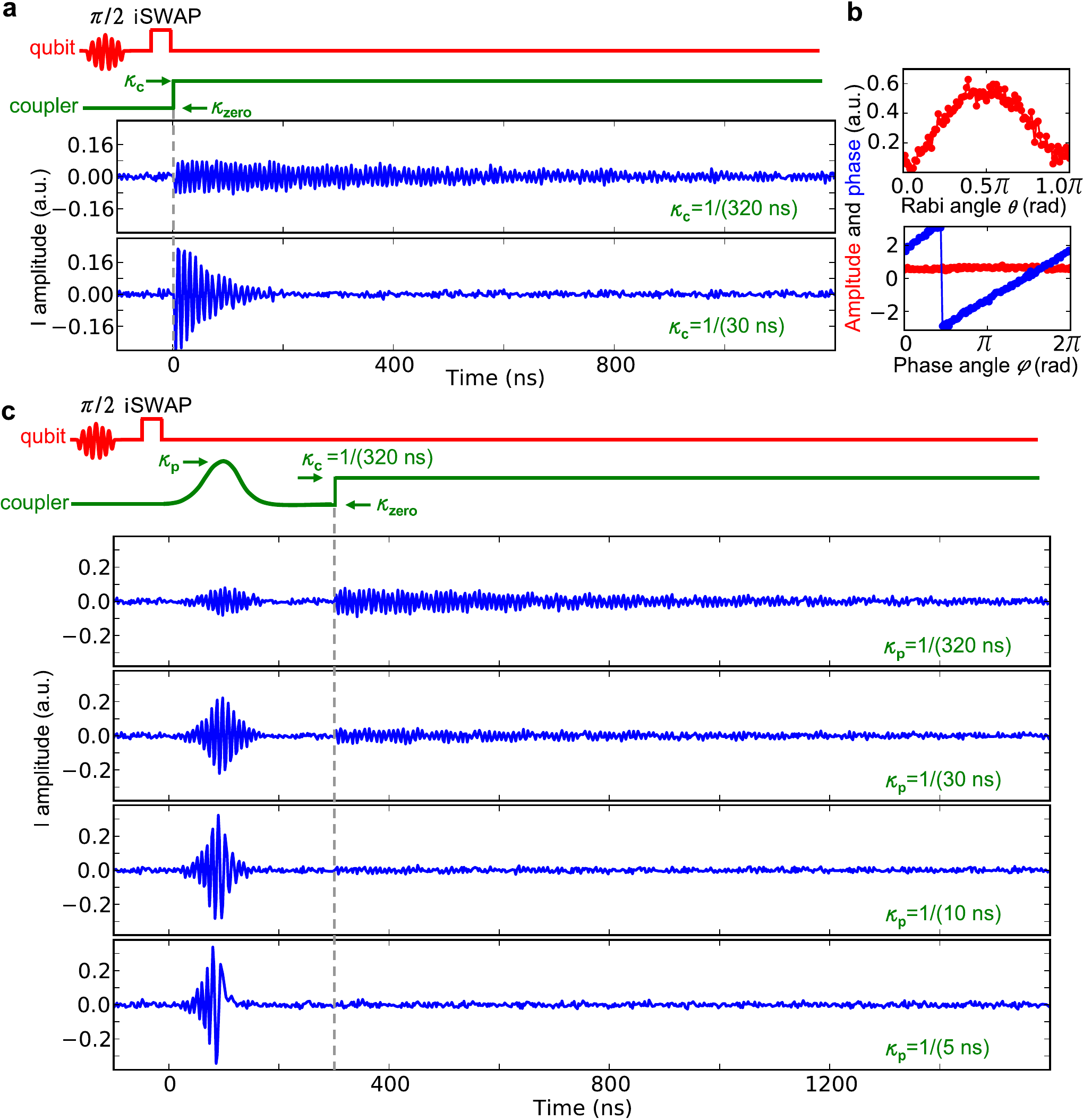}
\caption{Storage, release, and heterodyne detection of non-classical photon states.
\textbf{a}, Top: Superposition Fock state $(|0\rangle+|1\rangle)/\sqrt{2}$ is prepared with $\pi/2$ pulse to qubit, followed by iSWAP to resonator. Coupler is switched at $t=0$ from zero to either intermediate $\kappa_c \simeq 1/(320~{\rm ns})$ or larger $\kappa_c \simeq1/(30~{\rm ns})$, showing faster decay for larger $\kappa_c$. Middle and bottom: Heterodyne-detected $I$ quadrature for both $\kappa_c$ values. Mixer LO sideband frequency is 110 MHz, with $10^6$ averages.
\textbf{b}, Release of resonator superposition state $\cos(\theta/2)|0\rangle+e^{i\phi}\sin(\theta/2)|1\rangle$, with controlled Rabi angle $\theta$ and phase angle $\phi$. After state preparation, variable coupling changed from zero to $\kappa_c = 1/(320~{\rm ns})$, and released photons heterodyne-detected. Fourier transforms of $I$ and $Q$ quadratures yield signal amplitude and phase ($3.6 \times 10^5$ averages). Top: Amplitude for fixed $\phi=0$ has expected dependence on $\theta$, with only superposition states generating a signal (Fock states have completely undefined phase). Bottom: Signal for fixed $\theta = \pi/2$ as a function of $\phi$, with expected constant amplitude and phase scaling linearly with $\phi$.
\textbf{c}, First subpanel: Tailored, two-segment release pulse sequence for $(|0\rangle+|1\rangle)/\sqrt{2}$ superposition state, using a 200 ns Gaussian control pulse (100 ns FWHM, amplitude $\kappa_p$) followed by a rectangular pulse with $\kappa_c = 1/(320~{\rm ns})$, with intervening 100 ns delay. Second to fifth subpanels: Pulse amplitude $\kappa_p$ set to $1/(320~{\rm ns})$, $1/(30~{\rm ns})$, $1/(10~{\rm ns})$ and $1/(5~{\rm ns})$. $I$ quadrature amplitude depends on $\kappa_p$, with different amounts of energy released during Gaussian pulse; two top sub-panels show remainder released during final rectangular pulse. \label{fig.superpose}}
\end{center}
\end{figure}

These measurements confirm the phase-coherent capture and release of coherent states. To demonstrate that we can achieve the same control for non-classical states, we used the qubit to generate \cite{hofheinz:2009} the photon superposition state $(|0\rangle+|1\rangle)/\sqrt{2}$ and measured the release signal after turning on the coupling (Fig.\,\ref{fig.superpose}a). For an intermediate coupling $\kappa_c \simeq 1/(320~{\rm ns})$ and a large coupling $\kappa_c \simeq 1/(30~{\rm ns})$, the signal's exponential decay envelope has a time constant $T_d \approx 625$ ns and 69 ns, respectively, close to the expected $2/\kappa_c$, verifying that the coupling determines the release rate. The integrated energy for intermediate coupling is $7\%$ lower than for large coupling, attributed to greater intrinsic loss from the slower release.

We next tested the release and detection of the qubit-prepared superposition state $\cos(\theta/2) |0\rangle+ e^{i \phi} \sin(\theta/2) |1\rangle$, akin to previous work with static coupling \cite{boyzigit:2011, houck:2007}. The released photons were heterodyne-detected as a function of the Rabi angle $\theta$ and the phase angle $\phi$, with Fourier transforms of $I$ and $Q$ yielding the signal amplitude and phase. Figure\,\ref{fig.superpose}b shows the dependence of the signal amplitude on $\theta$, with $\phi=0$. The maximum amplitude is at $\theta = \pi/2$, corresponding to $(|0\rangle + |1\rangle)/\sqrt{2}$. The amplitude goes to zero for the pure Fock states at $\theta = 0$ and $\pi$ as expected, due to the loss of phase-coherence at the  Bloch sphere poles. When releasing the state $(|0\rangle+e^{i\phi}|1\rangle)/\sqrt{2}$ with $\theta = \pi/2$ and varying $\phi$, the signal has constant amplitude and phase increasing linearly with $\phi$.

The on-demand, real-time gating of the coupler enables precise shaping of the photon release waveform. Figure\,\ref{fig.superpose}c shows the tailored time-dependent release of the $(|0\rangle+|1\rangle)/\sqrt{2}$ photon state, modulating the coupling with a 200 ns Gaussian bias pulse with peak coupling $\kappa_p$, followed by a 100 ns delay and then completing the release with $\kappa_c = 1/(320~{\rm ns})$. Figure\,\ref{fig.superpose}c shows the $I$ quadrature signal for $\kappa_p = 1/(320~{\rm ns})$, $1/(30~{\rm ns})$, $1/(10~{\rm ns})$ and $1/(5~{\rm ns})$, with a Gaussian-like release waveform mimicking the coupler pulse. For the top three sub-panels, energy integrals show that $17.5\%$, $43.1\%$ and $100\%$ of the total stored energy is released during the pulse, with the remainder released after the $100$ ns delay. For couplings $\kappa_p \gtrsim 1/(10~{\rm ns})$, the release is completed during the Gaussian pulse. In contrast to fixed coupling, in which the waveform decays exponentially with time, the bottom sub-panel shows an exponential-like increase of the waveform, as needed for high-fidelity transfer of photonic information \cite{cirac:1997, korotkov:2011}.

We have demonstrated the phase-coherent, controlled capture and release of coherent and superposition photon states from a resonator, using a resonator-transmission line variable coupling. This powerful technique should allow long-range entanglement \cite{mariantoni:2011, wang:2011, neeley:2010,ritter:2012, duan:2001, briegel:1998}, where the shaped release we display in the last experiment is a key ingredient for high-fidelity state transfer \cite{cirac:1997, korotkov:2011}. This capability will further enable tunable coupling for resonator-based dispersive qubit readout, where time-domain control can minimize deleterious dephasing while maximizing measurement bandwidth and signal strength.

\noindent\textbf{Acknowledgements}
This work was supported by IARPA under ARO Award No. W911NF-08-01-0336 and
under ARO Award No. W911NF-09-1-0375. M.M. acknowledges support from an Elings
Postdoctoral Fellowship. R.B. acknowledges support from the Rubicon program of
the Netherlands Organization for Scientific Research. Devices were made at
the UC Santa Barbara Nanofabrication Facility, a part of the NSF funded
National Nanotechnology Infrastructure Network.

\noindent\textbf{Author Contributions}
Y.Y. designed and fabricated the samples, carried out the experiments and analyzed
the data. Y.Y. co-wrote the paper with J.M.M. and A.N.C., who also
supervised the project. Y.C. and D.S. developed the control infrastructure of
a custom-designed, FPGA-controlled ADC board. D.S. and P.J.J.O. provided assistance
with data-taking software. T.W. set up and calibrated the amplifier chain.
A.N.K. developed the theoretical model. All authors contributed to
the fabrication process, experimental set-up and manuscript revision.

\clearpage

%\clearpage

%\clearpage

%\clearpage

\end{document}

\setcounter{equation}{1}
\renewcommand\thefigure{S\arabic{figure}}

\title{Supplementary Information for ``Controlled catch and release of microwave photon states''}

\author{Yi Yin}
\affiliation{Department of Physics, University of California, Santa Barbara, CA 93106, USA}
\author{Yu Chen}
\affiliation{Department of Physics, University of California, Santa Barbara, CA 93106, USA}
\author{Daniel Sank}
\affiliation{Department of Physics, University of California, Santa Barbara, CA 93106, USA}
\author{P. J. J. O'Malley}
\affiliation{Department of Physics, University of California, Santa Barbara, CA 93106, USA}
\author{T. C. White}
\affiliation{Department of Physics, University of California, Santa Barbara, CA 93106, USA}
\author{R. Barends}
\affiliation{Department of Physics, University of California, Santa Barbara, CA 93106, USA}
\author{J. Kelly}
\affiliation{Department of Physics, University of California, Santa Barbara, CA 93106, USA}
\author{Erik Lucero}
\affiliation{Department of Physics, University of California, Santa Barbara, CA 93106, USA}
\author{Matteo Mariantoni}
\affiliation{Department of Physics, University of California, Santa Barbara, CA 93106, USA}
\affiliation{California NanoSystems Institute, University of California, Santa Barbara, CA 93106, USA}
\author{A. Megrant}
\affiliation{Department of Materials, University of California, Santa Barbara, CA 93106, USA}
\author{C. Neill}
\affiliation{Department of Physics, University of California, Santa Barbara, CA 93106, USA}
\author{A. Vainsencher}
\affiliation{Department of Physics, University of California, Santa Barbara, CA 93106, USA}
\author{J. Wenner}
\affiliation{Department of Physics, University of California, Santa Barbara, CA 93106, USA}
\author{Alexander N. Korotkov}
\affiliation{Department of Electrical Engineering, University of California, Riverside CA 92521, USA}
\author{A. N. Cleland}
\affiliation{Department of Physics, University of California, Santa Barbara, CA 93106, USA}
\affiliation{California NanoSystems Institute, University of California, Santa Barbara, CA 93106, USA}
\author{John M. Martinis}
\email{martinis@physics.ucsb.edu}
\affiliation{Department of Physics, University of California, Santa Barbara, CA 93106, USA}
\affiliation{California NanoSystems Institute, University of California, Santa Barbara, CA 93106, USA}

\maketitle

\clearpage

\section{Sample fabrication}
The device was fabricated using conventional photolithography and plasma etching on a multilayered structure. The resonator was made from a sputtered $150$ nm-thick superconducting aluminum base film on a sapphire substrate. The phase qubit and the superconducting quantum interference device (SQUID) in the variable
coupler were made using Al/AlO$_\text{x}$/Al Josephson junctions.  A low-loss dielectric, hydrogenated amorphous silicon, was used as the insulator in capacitors and wiring crossovers.

\section{Experimental setup and control electronics}
A detailed schematic of the resonator control and measurement system is shown in Fig.\,\ref{fig.supp1}a. The microwave excitation signal for the resonator is generated by mixing a digitally-synthesized intermediate frequency (IF) signal with a microwave frequency local oscillator (LO). The I and Q quadratures of the IF signal were generated using a field-programmable gate array (FPGA) and a customized 2-channel 14-bit digital-to analog converter (DAC). The RF output of the mixer was sent into the cryostat and routed through a circulator to the variable coupler and resonator. Microwave signals from the resonator were routed by the circulator to a cryogenic amplifier $G$ with 35 dB of gain and a noise temperature of 4 K. After further room-temperature amplification (60 dB of gain), the output signal was mixed down with the same local oscillator (LO) signal as the up-converter, generating the same IF (sideband) frequency. The $I$ and $Q$ quadratures were digitized at 500 megasamples/second using a $2$-channel $8$-bit analog-to-digital converter (ADC), and the signal passed to another FPGA for further processing. In the ``oscilloscope mode'', the digitized signals were sent directly to a computer without further processing, as shown in the time-trace data in Fig. 3 and Fig. 4 in the main text. In the ``demodulation mode'', the $I$ and $Q$ signals were multiplied by sine and cosine waveforms at the sideband frequency and summed in real time. The continuous summation signal guarantees rapid fast Fourier transform (FFT) processing once the data acquisition is complete; this was used for the data in Fig. 4b in the main text.

Qubit control used a system similar to that for the resonator. The microwave pulses for qubit control ($x/y$ rotations) were generated by mixing a continuous microwave signal and a shaped quasi-d.c. waveform from a $2$-channel $14$-bit digital-to-analog converter (DAC). Another quasi-d.c. pulse generator controls qubit $z$-axis rotations and measurement.

Figure 1 in the main text shows the circuit for the experiment. The variable coupler is connected a distance $d = 0.3$ mm from the grounded end of the $\lambda/4$ coplanar waveguide resonator ($4.8$ mm long). The coupler consists of a transformer with inductances $L_1$, $L_2$ and a negative mutual inductance $M$. The positive mutual inductance from the dc SQUID is $L_s=\Phi_0/4\pi I_c|\cos(\pi\Phi/\Phi_0)|$, where $I_c=1.6~\mu\text{A}$ is the critical current of the junction, $\Phi$ the applied magnetic flux and $\Phi_0=h/2e$ the magnetic flux quantum. The characteristic impedances of the microwave resonator and the transmission line are $Z_r=80~\Omega$ and $Z_0=50~\Omega$ respectively.

\section{Theoretical modeling of resonator $T_1$}
\begin{figure}[H]
\begin{center}
\includegraphics[width=10cm]{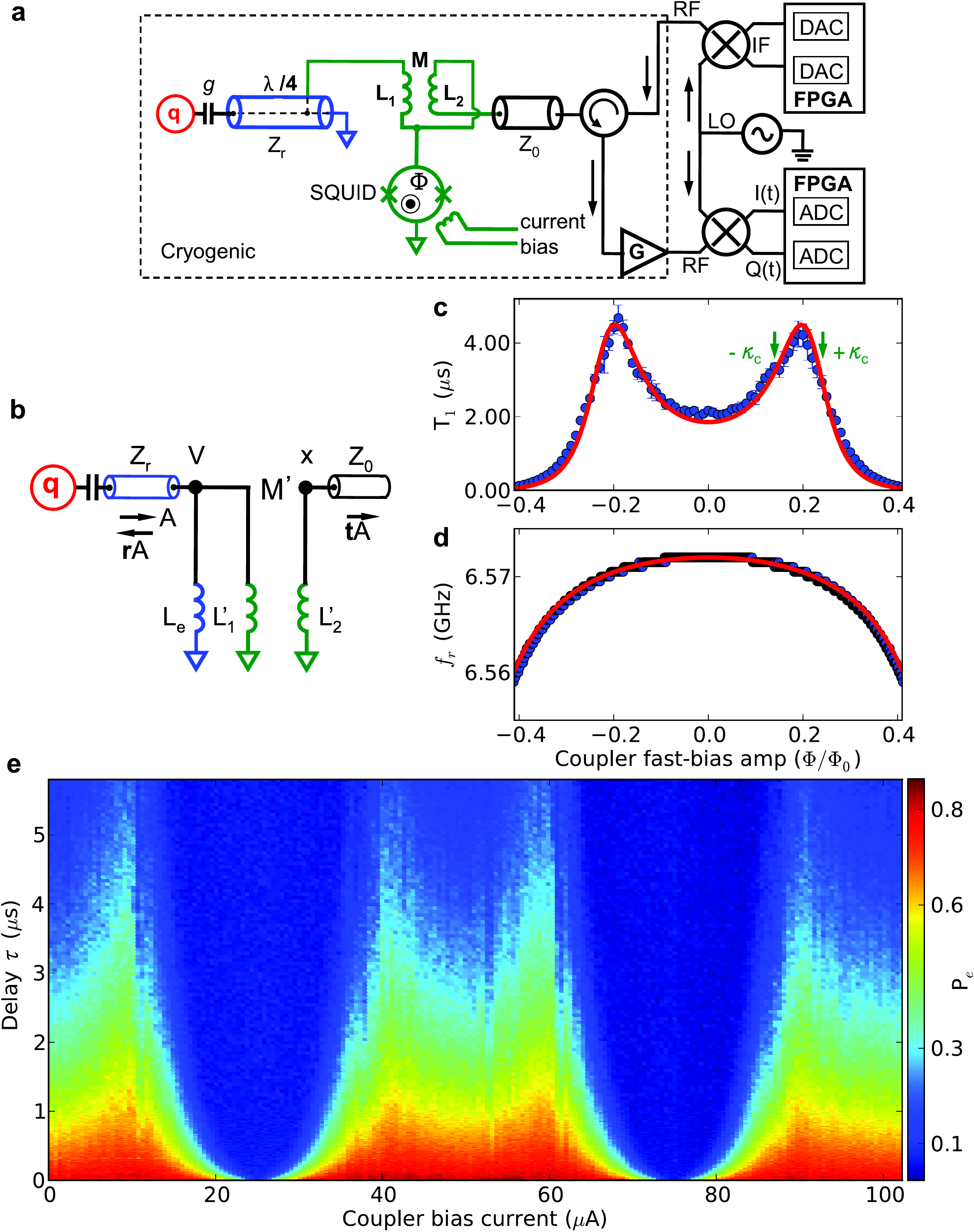}
\caption{Experimental setup and modeling of resonator lifetime $T_1$ and resonance frequency $f_r$.
\textbf{a}, Full schematic for resonator control and measurement electronics. Qubit $q$ is coupled with coupling strength $g$ to $\lambda/4$ resonator with characteristic impedance $Z_r$. Other end of resonator is grounded, with variable coupler connected a distance $\lambda/60$ from grounded end. Variable coupler comprises two inductances $L_1=L_2=480$ pH with a negative mutual inductance $M=-138$ pH, and a SQUID with tunable inductance $L_s(\Phi)$. Current bias to the flux line to the SQUID controls the SQUID inductance and thus the variable coupler. The coupler is connected to a transmission line with characteristic impedance $Z_0$, whose other end is connected through a circulator to a microwave excitation and measurement system (see text for details).
\textbf{b}, Simplified schematic diagram of the variable coupler end of the resonator. The section of the $\lambda/4$ resonator between the coupler and ground is approximated by an inductance $L_e$, and the transformer is replaced by two effective inductances $L'_1$, $L'_2$ with mutual inductance $M'$. When a voltage signal with amplitude $A$ travels from the left side of the resonator to the coupler, the wave is reflected from the coupler as $\textbf{r}A$ and transmitted in the transmission line as $\textbf{t}A$.
\textbf{c}, Blue dots are the experimental resonator lifetime $T_1$ extracted from the data shown in the main text in Fig.\,2b. Theoretical
evaluation of $T_1$ from Eq.\,(\ref{eq.rest1}) using the circuit design parameters is displayed as a red line. \textbf{d},  The resonator frequency $f_r$ as a function of coupler bias amplitude from spectroscopic measurements (blue dots) and compared with the theoretical prediction (Eq.\,\ref{eq.detune}). Arrows indicate coupling $\pm \kappa_c$ used for Wigner tomography in Fig.\,\ref{fig.supp2}, with the sign of $\kappa$ denoting polarity of the inductive coupling.
\textbf{e}, One-photon decay measurement, similar to Fig.\,2b in the main text, but using a slow coupler bias. The periodic response of the lifetime to the coupler current bias is evident. \label{fig.supp1}}
\end{center}
\end{figure}

Using an equivalent electrical circuit for the variable coupling experiment, shown in Fig.\,\ref{fig.supp1}c, we calculate the expected resonator $T_1$ due to coupling to the external $50~\Omega$ transmission line. We also calculate the small effect this coupling has on the resonance frequency of the resonator. Both calculations compare well with experiment.

We replace the short portion ($\sim \lambda/60$) of the resonator between the coupler connection and the resonator ground with an effective inductance $L_e$, and the transformer and coupling circuit with an equivalent $L_1'$, $L_2'$ and mutual inductance $M'$, with $L_2'$ connected to an infinite transmission line with characteristic impedance $Z_0$. We calculate the resonator reflection and transmission amplitudes $\textbf{r}$ and $\textbf{t}$.

The effective inductances $L_1'$, $L_2'$ and inductance $M'$ are given by
\begin{equation}
    L'_1=L_1+L_s,~L'_2=L_2+L_s,~M'=M+L_s.
\end{equation}
The coupler is turned off when $M'=0$. Note that because all the equivalent inductances include $L_s$, modulating $M'$ by changing $L_s$ modulates $L'_1$ and $L'_2$ as well:
\begin{equation}
    L'_1=(L_1-M)+M',~L'_2=(L_2-M)+M'.
\end{equation}

To calculate the inductance $L_e$, which represents the small length of resonator from the coupling point to ground, we impose a voltage $A e^{i\omega t}$ in the resonator traveling from the open (qubit) end towards the coupler end, and approximate the reflected voltage as $- A e^{i\omega t}$ (ignoring the portion transmitted into the transmission line). The voltage at distance $d$ from the grounded end is $V=Ae^{i\omega t}(e^{i\omega d/c}-e^{-i\omega d/c})=2iAe^{i\omega t}\text{sin}(\omega d/c)$, with $c$ the phase velocity of light in the resonator. The current at this point is $I = A e^{i\omega t}(e^{i\omega d/c}/Z_r+e^{-i\omega d/c}/Z_r) = 2 (A/Z_r)e^{i\omega t}\cos(\omega d/c)$, so
the wave impedance is $Z = V/I = i Z_r \tan(\omega d/c) = i \omega L_e$, yielding the effective inductance
\begin{equation}
    L_e=\frac{Z_r}{\omega} \, \tan \left (\frac{\omega d}{c} \right ) = \frac{Z_r}{\omega} \, \tan \left (\frac{2\pi d}{\lambda}\right )
\end{equation}
(note this is evaluated at $\omega = 2 \pi f_r$, the resonator frequency).

The calculation of the transmission and reflection amplitudes $\textbf{t}$ and $\textbf{r}$ is similar to the derivation in \cite{korotkov:2011}. Assume a voltage with amplitude $A$ is incident on the coupler from the left side of the resonator, with reflected voltage  $\textbf{r}A$ and voltage transmitted into the transmission line $\textbf{t}A$. The voltage across $L'_1$ is $V = (1+\textbf{r})A$, while the voltage across $L'_2$ is denoted by $x=\textbf{t}A$. The currents flowing into $L'_1$ and $L'_2$ are $I_1 = (1-\textbf{r})A/Z_r-V/(i\omega L_e)$ and $I_2 = -x/Z_0$, respectively. Using currents $I_1$ and $I_2$, we write equations for the voltage amplitudes $x$ and $V$:
\begin{align}
    x = i\omega M' \left [ \frac{(1-\textbf{r})A}{Z_r}-\frac{(1+\textbf{r})A}{i\omega L_e} \right ] -i\omega L'_2 \frac{x}{Z_0},\nonumber\\
    (1+\textbf{r})A = i\omega L'_1 \left [ \frac{(1-\textbf{r})A}{Z_r}-\frac{(1+\textbf{r})A}{i\omega L_e} \right ]-i\omega M'\frac{x}{Z_0}.
\end{align}
From these equations, we can calculate the reflection amplitude $\textbf{r}$ and transmission amplitude $\textbf{t}=x/A$ (note that $|\textbf{t}|^2 Z_r/Z_0+|\textbf{r}|^2=1$):
\begin{align}
    a \equiv & \frac{1+\textbf{r}}{1-\textbf{r}} = \frac{\frac{i\omega L'_1}{Z_r}+\frac{\omega ^2 M'^2}{Z_rZ_0(1+i\omega L'_2/Z_0)}}{1+\frac{L'_1}{L_e}-\frac{i\omega M'^2}{Z_0L_e(1+i\omega L'_2/Z_0)}},\nonumber\\
    \textbf{r} =& -\left (\frac{1-a}{1+a} \right ),\nonumber \\
    \textbf{t} =& i\frac{2\omega M'}{1+a} \left (\frac{1}{Z_r}+\frac{ia}{\omega L_e} \right ) \frac{1}{1+i\omega L'_2/Z_0}.
\end{align}
In the limit $\omega L_e \ll Z_r$ and $\omega M'\ll Z_0$, which apply here, the reflection and transmission amplitudes can be approximated as
\begin{align}
    \textbf{r} \approx & -1+2a \approx -1+i\frac{2\omega L_e L'_1}{Z_r(L'_1+L_e)},\nonumber\\
    \textbf{t} \approx & i \frac{2\omega L_e M'}{Z_r(L'_1+L_e)}\frac{1}{1+i\omega L'_2/Z_0}.
\end{align}

The decay time of the resonator is obtained from the transmission amplitude \cite{korotkov:2011}:
\begin{equation}\label{eq.rest1}
    1/\kappa_c = \frac{Q}{\omega}\approx \frac{\pi}{\omega |\textbf{t}|^2}\frac{Z_0}{Z_r}
    = \frac{\pi Z_r Z_0(L'_1+L_e)^2(1+\omega ^2 L'^2_2/Z_0^2)}{4\omega ^3 L_e^2M'^2},
\end{equation}
where $\omega = 2 \pi f_r$ is the resonator frequency.

The coupler bias dependence of the resonator $T_1$ is extracted from the data in the main text and shown in Fig.\,\ref{fig.supp1}c (blue dots). The predicted $T_1$ from Eq.\,(\ref{eq.rest1}), using the actual circuit parameters, is also displayed in Fig.\,\ref{fig.supp1}c (red line), in good agreement with the data.
We note that the inductive coupling changes sign when the coupler strength sweeps through zero coupling \cite{hime:2006, ploeg:2007}, verified by the expected $\pi$ phase change in a Wigner tomography measurement (see next section and Fig.\,\ref{fig.supp2}d).

The resonant frequency of the $\lambda/4$ resonator is primarily determined by the resonator length and characteristic impedance, but is also affected by the variable coupler. The change in resonance frequency with coupler bias can be measured experimentally, and verified by the following calculation: Compared to the frequency $f_r$ at zero coupling ($M' = 0$), the resonance frequency shifts by $\Delta f$,
\begin{equation}\label{eq.detune}
    \Delta f \approx -\frac{4{f_r}^2L^2_e M'}{\pi^2 Z_r(L_e+L_1-M)(L_e+L_1-M+M')}.
\end{equation}
We compare the coupler bias dependence of $f_r$, measured spectroscopically, with the frequency tuning from Eq.\,(\ref{eq.detune}), in Fig.\,\ref{fig.supp1}d. The frequency tunes over $\sim15$ MHz, a very small fraction of the resonator frequency. Previous experiments have demonstrated resonator frequency tuning using Josephson junctions or SQUIDs embedded in a resonator \cite{sandberg:2008, placios:2008}. Here the frequency tuning is quite small, and is a by-product of the variable coupler located outside the resonator.

We have measured the coupling dependence of the one-photon decay in Fig. 2b in the main text with a fast coupler bias. For comparison, a similar measurement was performed using the slow coupler bias, with data shown in Fig.\,\ref{fig.supp1}e. The bias range is expanded to show the periodic response of the lifetime to the coupler flux bias.

\section{Characterization of resonator states using swap spectroscopy and Wigner tomography}
\begin{figure}[H]
\begin{center}
\includegraphics[width=12cm]{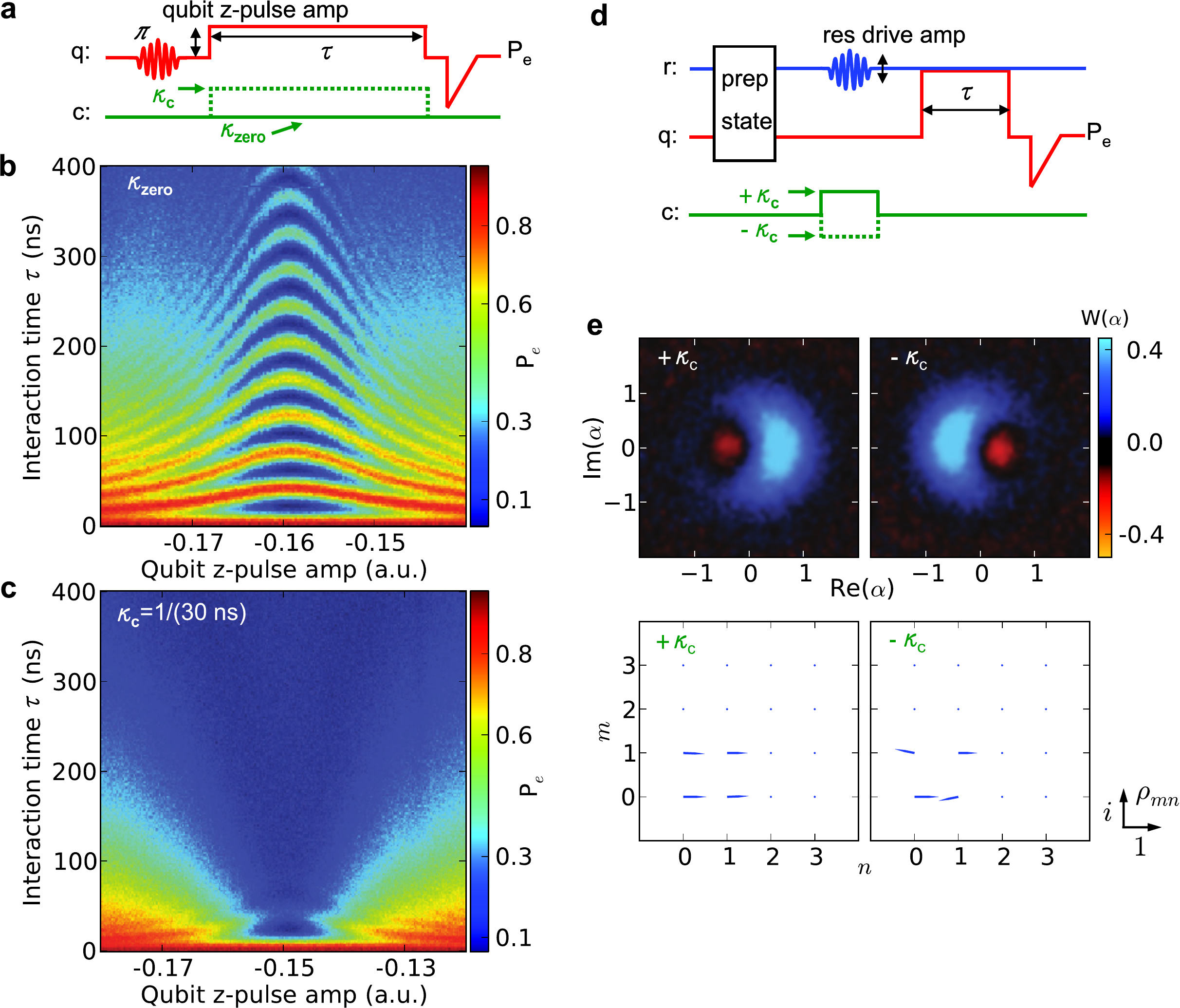}
\caption{Swap spectroscopy and Wigner tomography at different coupler biases.
\textbf{a}, Pulse sequence for swap spectroscopy at two coupler settings. The qubit is excited to $|e\rangle$ with a $\pi$ pulse and the excitation swapped to the resonator. The coupler is either left at zero coupling when the qubit is tuned on-resonance with the resonator, or switched immediately to $\kappa_c = 1/(30~{\rm ns})$. After a qubit-resonator interaction time $\tau$, the qubit excited state probability $P_e$ is measured.
\textbf{b}, Measured qubit probability $P_e$ for swap spectroscopy as a function of qubit $z$-pulse amplitude (detuning) and interaction time $\tau$, with the resonator coupling to transmission line set to zero. The chevron pattern representing qubit-resonator photon swapping is clearly visible.
\textbf{c}, In contrast, when the coupler is set to strong coupling $\kappa_c = 1/(30~{\rm ns})$, swap spectroscopy shows a rapid energy dissipation and a slight resonant frequency shift.
\textbf{d}, Pulse sequence for Wigner tomography. The box labeled ``prep state'' represents resonator preparation in the state $(|0\rangle + |1\rangle)/\sqrt{2}$. The coupler is set to zero coupling during the entire sequence except when the microwave source drives the resonator for the tomographic analyzer pulse. During the tomographic pulse, the coupler is set to the same coupling strength but with opposite sign $\pm \kappa_c = \pm 1/(2000~{\rm ns})$. Following the tomographic pulse the qubit is used to measure the resonator state.
\textbf{e}, Wigner functions $W(\alpha)$ (upper sub-panels) for the $(|0\rangle +|1\rangle)/\sqrt{2}$ resonator state, plotted as a function of the microwave tomography complex amplitude $\alpha$ in photon number units (51 by 51 pixels). We calculate density matrices (lower sub-panels) from each Wigner function. The negative sign for the coupling strength introduces a $\pi$ phase shift between the tomography pulse-induced state rotations, with a small phase error of $9.5^\circ$. \label{fig.supp2}}
\end{center}
\end{figure}

We measured the single photon lifetime $T_1$ to characterize the coupling strength, as discussed in the main text. We also used swap spectroscopy to perform an equivalent characterization, shown here with the coupler set to two representative coupling strengths. The pulse sequence is shown in Fig.\,\ref{fig.supp2}a, starting with the system initialized in the ground state. The de-tuned qubit was excited by a $\pi$ pulse to $|e\rangle$ and then tuned close to resonance with the resonator, using a qubit tuning $z$-pulse with variable amplitude. The variable coupler was either left at zero coupling ($\kappa_{\rm zero}$), or switched to a coupling $\kappa_c = 1/(30~{\rm ns})$ immediately after tuning the qubit. In either case, the coupling strength was fixed for the full qubit-resonator interaction time $\tau$. The qubit excitation probability $P_e$ was then measured using a triangular measurement pulse.

The qubit excited state probability $P_e$ is plotted versus the interaction time $\tau$ and the qubit $z$-pulse amplitude in Fig.\,\ref{fig.supp2}b and c, for weak and strong coupling, respectively. The chevron pattern due to the qubit-resonator photon swapping is evident in Fig.\,\ref{fig.supp2}b, from which we calibrate the iSWAP \cite{martinis:2009} pulse amplitude and duration. In contrast, the response in Fig.\,\ref{fig.supp2}c shows a rapid qubit-resonator relaxation, with energy strongly dissipated into the transmission line. We also note that the center of the chevron pattern in Fig.\,\ref{fig.supp2}c shifts in comparison to Fig.\,\ref{fig.supp2}b, due to the resonator frequency shift with coupler strength (Fig.\,\ref{fig.supp1}d).

The coupling strength changes sign when the coupler bias sweeps through the zero coupling point. An indirect phase-sensitive method, Wigner tomography \cite{hofheinz:2009}, was used to detect this coupling sign change. The pulse sequence is shown in Fig.\,\ref{fig.supp2}d. The resonator was prepared in the superposition $(|0\rangle+|1\rangle)/\sqrt{2}$ state, and then driven by a variable-amplitude classical Gaussian microwave tomography pulse. The coupler was set to two different values during the tomography pulse, such that the coupling strength $\pm \kappa_c$ had the same amplitude but opposite signs. The microwave tomography pulse, passing through the coupler, displaces and rotates the resonator state in the resonator phase space; the opposite coupling signs give opposite rotation directions to the resonator state for the same tomography pulse. The qubit was then tuned on-resonance with the resonator for a variable time $\tau$, after which the qubit state was measured. Measurements of the qubit excitation probability $P_e(\tau)$ were analyzed to yield the Fock state probability $P_n(\alpha)$, where $\alpha$ is the complex amplitude and phase of the tomography pulse. The Wigner quasi-probability distribution $W(\alpha)$ was calculated by evaluating the parity $W(\alpha) = \sum (-1)^n P_n(\alpha)$.

The Wigner functions measured with the two signs of coupling strength $\pm \kappa_c$ are shown in Fig.\,\ref{fig.supp2}e. The Wigner functions clearly show a relative rotation angle of $\sim \pi$. The density matrices of the resonator states can be calculated from the Wigner functions and projected onto the number basis $\rho_{mn}=\langle m|\rho|n\rangle$, shown in the lower sub-panel of Fig.\,\ref{fig.supp2}e. Here, we represent each element in the density matrix by an arrow, whose length and direction correspond to the magnitude and phase of $\rho_{mn}$. The direction of the arrow for $\rho_{01}$ ($\rho_{10}$) contains the relevant phase information, showing a $\approx\pi$ phase shift with a small phase error of $9.5^\circ$.

\section{Calibration of trapped coherent state photons}

When a microwave Gaussian pulse with amplitude $\alpha$ and duration $t_d$ is used to create a coherent photon state in the resonator, the state can be probed with a qubit through an on-resonance interaction, by measuring the qubit excited state probability $P_e(\tau)$ as a function of the interaction time $\tau$. The photon state probability distribution $P_n(\alpha)$ can be resolved in the Fock number basis $|n\rangle$ by decomposing $P_e(\tau)$ into its discrete Fourier components $f_n = n f_1$, where $f_1 = g/\pi$ is the vacuum Rabi frequency \cite{hofheinz:2009}.

\begin{figure}[H]
\begin{center}
\includegraphics[width=12cm]{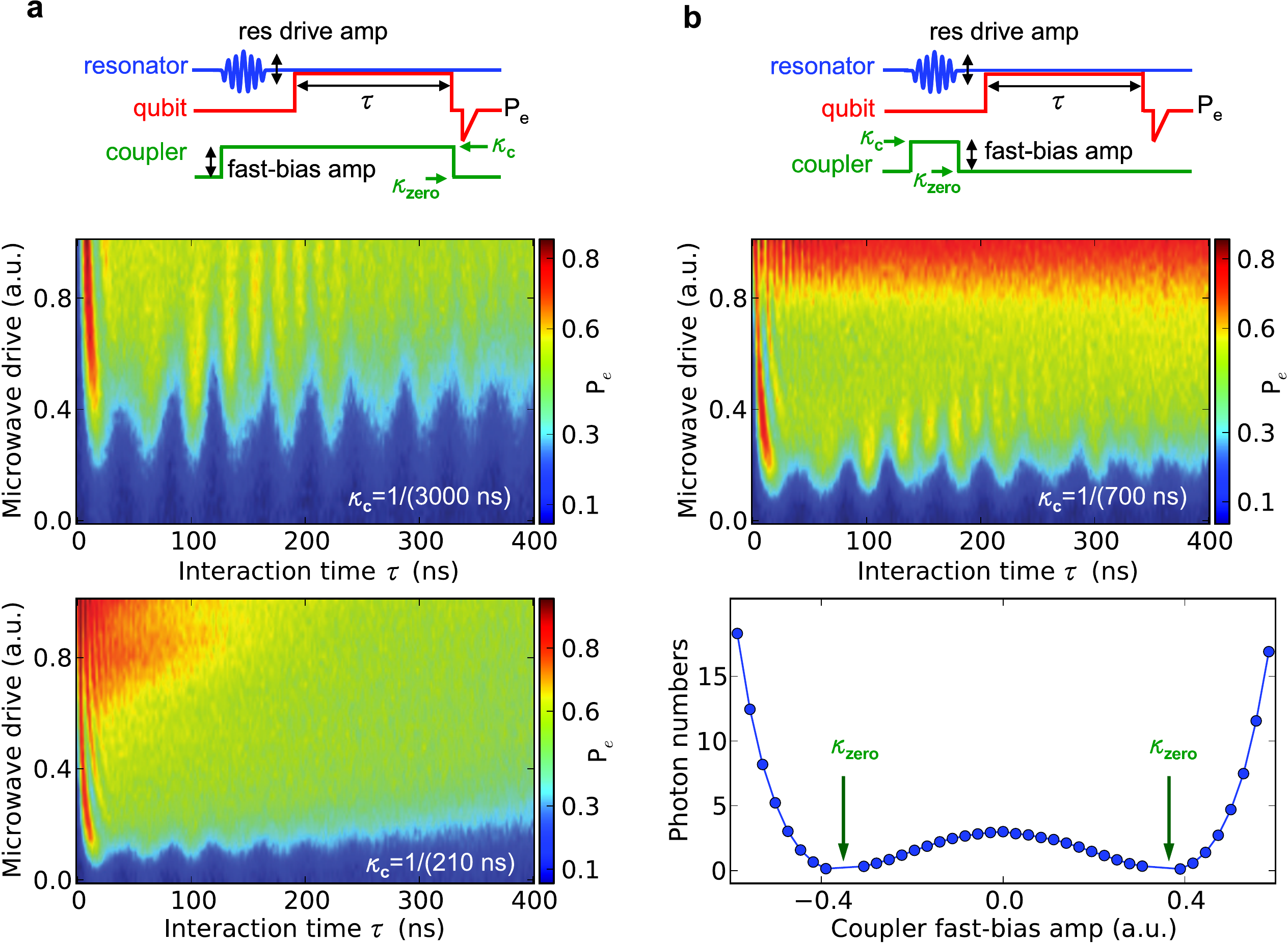}
\caption{
Calibration of coherent state generation for different coupler strengths.
\textbf{a}, Top sub-panel: Pulse sequence to generate a resonator coherent state and then perform a qubit measurement. This measurement was performed for two different coupling strengths, $\kappa_c \simeq 1/(3000~{\rm ns})$ and $\kappa_c \simeq 1/(210~{\rm ns})$, with the coupler set to this value prior to the excitation pulse and left at this value during the qubit-resonator interaction. Data in bottom two panels show the qubit excited state probability $P_e$ versus interaction time $\tau$ and microwave drive amplitude. The Gaussian microwave pulse was 12 ns in duration (6 ns FWHM) for both panels a and b.
\textbf{b}, Top sub-panel: Pulse sequence to generate a resonator coherent state with a coupling strength $\kappa_c \simeq 1/(700~{\rm ns})$, with the coupling strength set to zero during the subsequent qubit-resonator interaction. Middle panel shows the qubit $P_e$ as a function of interaction time $\tau$ and microwave drive amplitude. A population analysis yields the average trapped photon number $\langle n \rangle$ for different coupling strengths during the microwave drive pulse, shown in the bottom sub-panel, for a microwave drive amplitude $\alpha = 1.0$, the same as $1.0$ in the vertical axis of the middle sub-panel. Coupler biases yielding zero average photon number are marked by arrows. \label{fig.supp3}}
\end{center}
\end{figure}

We used this measurement to calibrate the coherent state stored in the resonator for different coupling strengths and different microwave drive amplitudes $\alpha$, with a fixed duration. In Fig.\,\ref{fig.supp3}a we set the coupler strength to one of two values $\kappa_c \simeq 1/(3000~{\rm ns})$ and $\kappa_c \simeq 1/(210~{\rm ns})$, after which we excited the resonator with a variable amplitude Gaussian microwave pulse, and measured the qubit after a qubit-resonator interaction time $\tau$. The top sub-panel shows the pulse sequence, the middle sub-panel the qubit excitation probability $P_e$ as a function of the microwave pulse amplitude and interaction time $\tau$ for the smaller coupler strength, and the bottom sub-panel the same measurement for the larger coupler strength. A horizontal line cut (not displayed) shows a periodic but low amplitude oscillation for small drive amplitude $\alpha$, transforming to a clear ringing-collapse-revival pattern for larger $\alpha$. When the coupler is set to a small coupling (middle sub-panel of Fig.\,\ref{fig.supp3}a), it is hard for the microwave source to excite the resonator but the resonator has a large $T_1$ for trapped photons. When the coupler is instead set to a large coupling (bottom sub-panel of Fig.\,\ref{fig.supp3}a), photons enter the resonator easily yielding a larger excitation amplitude, but the lifetime is shorter, illustrated by the rapid decay for larger $\tau$.

To achieve both long photon lifetimes and low-power excitation, we instead set the coupler to a large value during the microwave drive pulse, then set the coupling to zero to trap the photons during the qubit measurement (Fig.\,\ref{fig.supp3}b). A representative qubit-resonator interaction measurement for a coherent resonator state is shown in the middle sub-panel of Fig.\,\ref{fig.supp3}b, with the coupler set to $\kappa_c \simeq 1/(700~{\rm ns})$ during the microwave excitation pulse. We performed a series of measurements with varying coupling strengths, which were analyzed to give the photon distribution $P_n(\alpha)$. For a fixed microwave pulse amplitude ($\alpha = 1.0$ in the vertical axis of Fig.\,\ref{fig.supp3}b), we calculated the average photon number $\langle n(\alpha) \rangle = \sum_m n P_n(\alpha)$. We display $\langle n \rangle$ as a function of coupler drive amplitude in the bottom sub-panel of Fig.\,\ref{fig.supp3}b; this is the calibration method used for the experiment shown in Fig.\,3 in the main text.

\clearpage

\clearpage

\clearpage

